\documentstyle[12pt,aps,preprint,tighten,epsf]{revtex}
\begin{title}
{\bf\large Chaotic Behavior of a One-dimensional Model Atom in an
Intense Field
\thanks{
This work was supported in part by the National Natural Science Foundation of China/19674011,
the Science Foundation of the CAEP,
National High Technology Committee of Laser and the Hong Kong
Baptist University Faculty Research Grant.
}
}

\end{title}
\author{ Liu Jie $\,^{1,2}$, \,\,\,\,\,\, Chen Shi-Gang  $\,^1$ \,\,\,\,\,\, and  Bambi  Hu $\,^{2,3}$ \,\,\,\,\,\,\\
1.Institute of Applied Physics and
Computational Mathematics,\\ 100088 Beijing, China\\
2.Department of Physics and Centre for Nonlinear and Complex Systems
,\\ Hong Kong Baptist University, Hong Kong\\
3.Department of Physics, University of Houston, Houston,TX 77204 USA}

\begin{document}
\date{Received 4 July 1997}
\maketitle
\baselineskip=1.0\baselineskip
{ \it  In this paper we describe the rescattering process
in optical field ionization
through a one-dimensional  model, which improves  the well-known quasistatic model
by adding the smoothed Coulomb potential in its second step.
The
above-threshold
ionization spectra and high-order harmonic generation are calculated
from this model.
They are qualitatively in agreement with the quantum results and experiments.
In particular,
we find that this
 model is characterized by chaotic scattering. Modern nonlinear
theory is used to analyze this dynamical  system.
It is found that singular
 self-similar fractal structure exists in the phase-dependence energy spectra,
   and the unstable
manifolds constitute the chaotic scattering pattern.
Our results also demonstrate  a close connection of 
irregular trajectories with the energy spectra and  high-order harmonic generation.
We conclude that the chaotic behavior plays an important role in
 this one-dimensional
model of  optical field ionization.
  }
\\PACC:
\newpage
\section{INTRODUCTION}
In recent years there have been significant advances in high-intensity
laser technology. One major use of these laser systems has been for
studies of the response of atoms and molecules to such intense fields.
It has led to the discovery of a whole range of nonperturbative
phenomena, such as, multiphoton ionization (MTI), above-threshold
ionization (ATI), and high-order harmonic generation (HOHG).
Because  the fields are very intense, the traditional perturbation 
expansion of
the wave function in terms of the field-free states will fail completely 
to describe the dynamics of the system.
Therefore, these problems are challenging theoretically and 
 have recently received much attention both experimentally and theoretically
.$^{[1,2]}$

 It has been concluded by numerical calculation of the one-electron
 Schrodinger equation that these phenomena can be understood in the
 context of single-electron ionization dynamics.
 Various nonperturbative theories are developed to explain ATI and HOHG
 phenomena.
  Among  these theories
  the quasi-static model is commonly used in treatment
 of the optical field ionization (OFI) problem  in the regime of low
 field  frequency.$^{[3,4]}$
 In this  model, an electron can be
 considered to acquire its energy   in a two-step process: first
 the electron is removed from the atom, thereby overcoming the ionization 
potential, and then it interacts with the laser field.
 The ionization probability of the
atom in the first step is usually described by the tunneling theory.
Since it is in the continuum state, the evolution of an electron
wave packet after tunneling is usually described by classical
motions governed only by the laser field.
This simple two-step quasi-classical model  accounts for
many of the detailed characteristics of the photon and electron
emission process and provides significant insights into the excitation
dynamics under intense field conditions.

However, in the case of an intense linearly polarized laser,
the amplitude of the quiver motion of an electron after tunneling
is large.
Then
 the probability that an ionized electron returns
 to the vicinity of the nucleus and is rescattered by the ion is not
 negligible.
 This problem attracts much attention and becomes an active point
 recently.
 Many classical and quantum theories are developed  to treat this rescattering
 process.$^{[5,6,7]}$ It is found that this process may increase the fraction of the
 electron with higher energy , and is responsible for other important
 phenomena , such as the 'cut off' law in high-order harmonic generation.
 Different from the above works, in this paper
 we try to consider the rescattering process in the frame of
 one-dimensional (1D) quasiclassical
 model, which is derived from
 natural extending of the original quasistatic model by
 adding a smoothed Coulomb  potential
  in its acceleration step .
 In sec.II  we will introduce this 1D model. According to this
 model the above-threshold energy spectra and high-order harmonic generation
 in the rescattering process are calculated in sec.V.
 Our results show that the rescattering process increases the higher energy
 electron and generates high-order harmonics. Additionally, the 'cut off' law is
 identified by this model.
 In particular, based on nonlinear theories, we discover a new
 physical process - chaotic scattering underlying optical field ionization
  and give a complete dynamical explanation for them (secs.III and IV).
We find that  self-similar and fractal  structure exists in the plots of
ATI energy vs initial field phase. We  also identify those
unresolved regions of ATI energy with chaotic layers. The location and
width of those singular regions are estimated by a nonlinear theory.
In sec.VI we analyze the dynamical structure in
Poincare section which causes the chaotic scattering.
Finally
in sec.VII we summarize our main results and discuss some problems
for future study.
\section*{II. The Model}
First let us recall the well-known quasistatic theory briefly.
As $\gamma \equiv (I_0/2U_p)^{1/2}\ll 1, \hbar \omega /2U_p\ll 1$ and field
strength $F < F_{th}\equiv Z^3/16n_{eff}^4 a.u.$ ( atomic units are used in this paper),  tunneling
ionization occurs. Here, $I_0$ is the ionization potential of the electron,
 $U_p=\frac{Z^2e^2F^2}{4m_e\omega^2}$
is the ponderomotive potential, $F_{th}$ is the threshold field,  $Z$ is 
the core charge and $n_{eff}$ is the effective principle quantum number. 
The corresponding threshold intensity of $F_{th}$ is
\begin{equation}
I_{th}=1.37\times 10^{14}(I_0/13.6eV)^4/Z^2\qquad (W/cm^2) . 
\end{equation}
Based on the Landau tunneling ionization theory,$^{[8]}$,
 Ammosov, et al.$^{[9]}$
suggested a simple formula for the ionization rate of complex atoms:
\begin{equation}
\Gamma(F)=1.61\frac{Z^2}{n_{eff}^{4.5}}(\frac{10.87Z^3}{n_{eff}^4F}%
)^{2n_{eff}-1.5}\exp (-\frac 23\frac{Z^3}{n_{eff}^3F})\qquad a.u. 
\end{equation}
This formula  is commonly used in estimating the ATI energy of the ionized 
electrons.$^{[10]}$

The second part of the quasistatic procedure uses classical mechanics
to describe the evolution of an electron packet. After tunneling, the
electron motion in the field is given by

$$ x = x_0 [cos(\omega t)] + v_{0x} t + x_{0x} , $$
  \begin{equation}
 v_x = -v_0 sin(\omega t) + v_{0x} ,
\end{equation}
where $v_0=ZeF/m_e\omega, x_0=ZeF/m_e\omega^2$.
The energy associated with the velocity $v_{0x}$  constitutes
the ATI energy and can be evaluated by averaging over the external field
\begin{equation}
E_{ATI} = U_p[1+2\sin^2\phi_0] ,
\end{equation}
where $\phi_0 = \omega t_0$ denotes the field phase.

In a linearly polarized intense  laser  field , the amplitude of the quiver motion is
large. Then the ionized electrons are driven by the laser field
 to return to the vicinity
of the ions and are rescattered by them. This process will result in many
important phenomena, such as 'plateau' in ATI peaks, 'ring' in angular
distribution and 'cut-off'
law of high-order harmonic generation. To take the rescattering process
into consideration, we improve the above model by adding a smoothed 
Coulomb potential in its second procedure.
Then the governing equation is (the atomic units is used for simplicity),
\begin{equation}
H = \frac{p^2}{2} -\frac{1}{\sqrt{x^2+\alpha}} + e x F cos(\omega t) ,
\end{equation}
where  $\alpha = 0.1$ the 'softening' factor to normalize the Coulomb potential.

The initial position of an electron  borne outside the
potential barrier  at time $t_0$ is determined by the tunneling ionization
theory,
\begin{equation}
U_{eff}=-\frac{Ze^2}{\sqrt{x_0^2+\alpha}} + e F x_0 cos(\omega t_0) = I_0 , 
\end{equation}
where $I_0=0.5 $ is the ionization potential.

Take $\pi/2<\omega t_0<3\pi/2$ as an example. It is observed that
the electron corresponding to the maximum field strength ($\omega t_0 =
\pi$) is closest to the nucleus. As $\omega t_0 =\frac{\pi}{2}$ or
$\frac{3\pi}{2}$, the initial electron position tends to  positive
infinity.
Due to the fact that an electron will be in the continuum state after
 tunneling ionization,
we prefer to choose the initial energy rather than its initial momentum of an ionized electron born at time $t_0$
to be zero. That is,
if the external field is removed away at time $t_0$
 the ionized electron  will
escape away from the Coulomb potential.  On the contrary,
the condition that the initial
momentum is zero cannot guarantee the escaping of the ionized
electron.$^{[11]}$
Then the initial momentum of an ionized electron can be derived
  from the above assumption.
  These initial conditions
 are sufficient for us to study the rescattering effects in the following sections.

\section*{III. Self-Similar  Fractal Structure and Scaling Law}
It is convenient to introduce the compensated energy $E_c$ advocated by
Leopold and Percival,$^{[12]}$
\begin{equation}
E_c = [v_x-(F/\omega)sin(\omega t)]^2/2-1/\sqrt{x^2+\alpha}  ,
\end{equation}
When an electron is ionized completely and cannot return to the
nucleus again, the Coulomb potential is weak enough and $E_c$ tends to a
positive constant value which is just the ATI energy ($E_{ATI}$) in
ultrashort pulse laser.
\begin{figure}
\epsfxsize=8cm
\epsffile{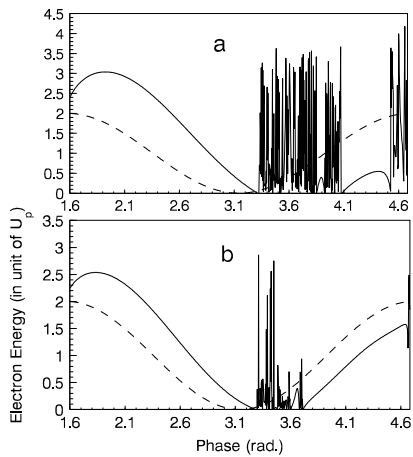}
\smallskip
\caption{
Plots of electron energy  vs field phase. Dashed lines and
solid lines denote the results from original model
and the  present model respectively.
a) $F=0.06 a.u. ,\omega=0.07 a.u. \,\,$b)$F=0.06 a.u.,\omega=0.04242 a.u.$.
 }
\end{figure}

\begin{figure}
\epsfxsize=8cm
\epsffile{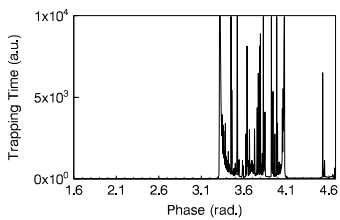}
\smallskip
\caption{
Trapping time vs field phase. $F=0.06 a.u.,\omega=0.04242 a.u.$.
 }
\end{figure}

In Fig.1 we demonstrate the phase dependence of the ATI energy.
A Runge-Kutta algorithm of the $4{th}$ order in $x$ and $p$
is employed in the calculations.
One thousand  points distributed equally in the range $[\pi/2,3\pi/2]$ are chosen as the
initial phase. It was observed that
the curves calculated from  the original model where the Coulomb potential is
neglected are smooth over all the range; however, in our new
model things are much different.
The dependence of $E_{ATI}$ on the initial phases is poorly resolved in the
region $[3.32,4.05], [4.52,4.71]$ for $\omega=0.07 a.u.$ and in the region
$[3.30,3.70], [4.69,4.71]$ for $\omega=0.04242 a.u $. Magnification of the
unresolved regions shows that an arbitrarily small change in the initial phase
may result in a large change in the final electron energy. We define
the escaping time of an electron as the moment $T$ $^{[13]}$ satisfying 
\begin{equation}
E_c(t=T) =0 ,\,\,\,\, E_c(t>T) > 0 .
\end{equation}
Naturally,  the trapping time of an electron
is defined as $\Delta T= T-t_0$ . A comparison between
Fig.1 and Fig.2 shows that the unresolved regions in Fig.1 coincide
 with the singularities of the trapping time.

 We call a value $\phi_s$ a singularity if, in any small
 neighborhood domain
 there is a pair of $\phi$  which have different signs in
 their final moments. These singular values relate to the
 infinite trapping  time .  Generally in phase space they constitute a
 nonattracting  hyperbolic
 invariant set.  This invariant set has a fractal structure. Figure 3  shows
 successive magnification of the unresolved region and demonstrates clearly
 the existence of self-similarity.
\begin{figure}
\epsfxsize=8cm
\epsffile{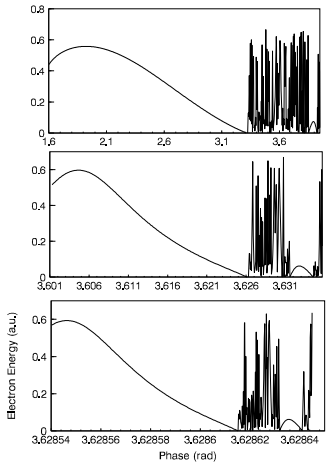}
\smallskip
\caption{
Successive magnification of the plot for $F=0.06 a.u., \omega=0.07 a.u.$.
 }
\end{figure}

 To calculate the fractal dimension of the singular points $\phi_s$, we
 can employ the uncertainty exponent technique to obtain the fractal
 dimension of the singular set.$^{[14]}$ We randomly choose many values of $\phi$
 in an interval containing the fractal set. We then perturb each value
 by an amount $\epsilon$ and determine whether the final momentum
  corresponding
 to initial $\phi, \phi-\epsilon$ and $\phi+\epsilon$ have the same sign.
 If so , we say that the $\phi$ value  is $\epsilon$-certain; if not,
 we say it is $\epsilon$-uncertain. We do this for several $\epsilon$
 and plot on a log-log scale the fraction of uncertain $\phi$ values
 $f(\epsilon)$. The result is plotted in Fig.4 which shows a good
 straight line and indicates a power law
 dependence $f(\epsilon) \sim \epsilon^\gamma$,
 where $\gamma = 0.15$.
 The exponent $ \gamma $ is related to the dimension of the fractal
 set of the singular $\phi$ values by
 \begin{equation}
 D_0 = 1 - \gamma=0.85 .
 \end{equation}

 In another aspect, it is interesting to make an intensive investigation on the
 escaping process of an electron after tunneling. We find that
 the trajectories demonstrate much different behavior for different
 initial phase. Some electrons drifts away from nucleus without a collision,
 others can undergo multiple collisions. In Fig.5 we plot the probability
 vs. the collision times by making a statistics on 4000 trajectories
  corresponding to different phases. It is found that, approximately
   the probability decays
  with increasing  collision times and satisfies the following
  scaling law
  \begin{equation}
  P_{col} \sim e^{-\kappa N} ,
  \end{equation}
   where $\kappa = 0.21$ , N denotes the collision number.

  Obviously, the possibility  of multiple collisions
   decreases exponentially
  with increasing  collision number. Although chaos implies
  the existence of  infinite times of collisions, its probability
  is zero. This is because  the hyperbolic invariant set corresponding to
  the trapped trajectories has zero measure.

\begin{figure}
\epsfxsize=8cm
\epsffile{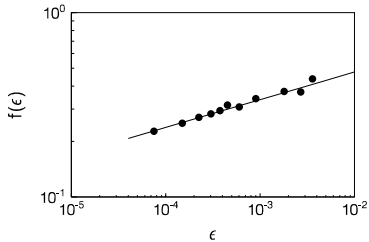}
\smallskip
\caption{
Calculation of fractal dimension
in the case $F=0.06 a.u., \omega=0.07 a.u.$
 }
\end{figure}

\begin{figure}
\epsfxsize=8cm
\epsffile{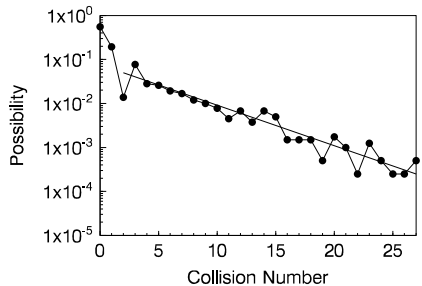}
\smallskip
\caption{
Probability of multiple collisions obtained from making statistics on
4000 trajectories corresponding to different initial phase (circle). The
straight line denotes the scaling law mentioned in the text.
 }
\end{figure}
\section*{IV. Unresolved Regions Resulting From Chaotic Layers}
In  previous sections we discovered that the unresolved regions have an
irregular, self-similar fractal structure.  In this section
we will discuss the dynamical source of these phenomena. We will
determine the location and width of the unresolved regions by using
modern nonlinear theory.

First let us review some important conclusions $^{[15]}$ on a two-dimensional (2D)
area-preserving mapping described by
\begin{equation}
\bar E = E + \Delta E,\,\,\,\,\bar\phi = \phi + f(\bar E)
\end{equation}
In the neighborhood of a separatrix $(E_s)$, where the period
$f(E_s)$  tends to infinity, even small changes in energy $E$ may
result in considerable changes in phase $\phi$. This is the
very cause of local stochasticity. The following expression can be
served as a good evaluation of the stochastic region
boundary
\begin{equation}
|\frac{\delta\bar\phi}{\delta\phi} - 1| > \pi .
\end{equation}
In the rest of this section we shall
employ the Kramers-Henenberger (or acceleration) gauge  and  
atomic units will be used to simplify our
formula.
The variable transformations are
\begin{equation}
P = p + \frac{\epsilon}{\omega} sin \omega t ,\,\,\,\,\,
X = x - \frac{\epsilon}{\omega^2} cos \omega t .
\end{equation}
The new Hamiltonian is
\begin{equation}
K(X,P) = \frac{P^2}{2} - \frac{1}{\sqrt{(X+\frac{\epsilon}{\omega^2}
cos\omega t)^2 + \alpha}} .
\end{equation}
The corresponding equations of motion are
\begin{equation}
\dot X = P ,\,\,\,\,
\dot P = -\frac{X+\frac{\epsilon}{\omega^2} cos\omega t}
 {(X+\frac{\epsilon}{\omega^2}cos\omega t)^2 + \alpha)^{\frac 3 2}} .
 \end{equation}
The fundamental property of the Hamiltonian system is that the
 Cartan differential 2-form
\begin{equation}
dP\wedge dX-dH\wedge dt
\end{equation}
is preserved under Hamltonian flow.
As the term  $\frac{\epsilon}{\omega^2}cos\omega t$ in (16) vanishes , the energy
\begin{equation}
E_0 = \frac{P^2}{2}-\frac{1}
{\sqrt{X^2+\alpha}} .
\end{equation}
 is an integral constant. Otherwise,
its evolution is time-dependent and can be described approximately
by virtue of Eqs. (16) and (18) 
\begin{equation}
\frac{dE_0}{dt} = \frac{2\epsilon}{\omega^2} PXcos\omega t/(X^2+\alpha)^2 .
\end{equation}
To   evaluate   the jump in energy, we consider the following straight-line
 trajectory as the zero${th}$-order approximation,
 \begin{equation}
 P = v_c ,\,\,\,\,\,
 X = v_c (t-\tau),
 \end{equation}
 where $v_c$ and $\tau$ are the momentum and  time, respectively , at $X = 0$.
 Then the jump of the  energy  for each return is obtained by integrating
 Eq.(19) from negative infinity to positive infinity,
 \begin{equation}
 \Delta E_0 = \frac{2\epsilon}{\omega^2}\int_{-\infty}^{+\infty}
 v_c^2(t-\tau) cos\omega t/(v_c^2(t-\tau)^2+\alpha)^2 dt .
 \end{equation}
 The above integral can be rewritten in a  simpler form,
 \begin{equation}
 \Delta E_0 = \frac{\epsilon}{v_c^2}sin\omega\tau
 \int_{-\infty}^{+\infty}
 \frac{cos (s)} {s^2+\alpha \omega^2/v_c^2} ds ,
 \end{equation}
 The integral included in the above formula can be expressed in term of a  
 Bessel function,
 \begin{equation}
 K_{\frac 1 2}(z) = \frac{(2z)^{\frac 1 2}}{2\sqrt{\pi}}
 \int_{-\infty}^{+\infty}\frac{cos t}{(t^2+z^2)} dt ,
 \end{equation}
 where $z=\alpha^{\frac 1 2}\omega/v_c$
Now we divide our problem into the following two cases.

A. Case 1, $E_{s1} =0$. 
The jump of energy is rewritten as
\begin{equation}
\Delta E_0 = \delta E_{s1} sin\phi,\,\,\,\,\phi=\omega \tau ,
\end{equation}
where
\begin{equation}
\delta E_{s1} = \frac{\epsilon}{v_{c1}^2}
 \int_{-\infty}^{+\infty}
 \frac{cos(s)} {s^2+\alpha \omega^2/v_{c1}^2} ds .
 \end{equation}
 and $v_{c1}$ is derived from Eq.(18) by setting $E_0=E_{s1}$ and  $X=0$.
As the 'softening' factor $\alpha$ is usually chosen as a small value,
the Keplerian Law  of the two-body problem is available approximately.
Therefore, we obtain the corresponding change of phase as follows,
\begin{equation}
\bar\phi = \phi + \frac {\omega \pi}{|E_0|^{\frac 3 2}} .
\end{equation}
Equations (24) and (26)  construct a 2D mapping
which reflects the dynamical properties of the system (15) in the vicinity
of $E=E_{s1}$. The
area-preserving nature of this mapping, a consequence
of the invariance of the Cartan differential 2-form (17),  can be checked easily.

In terms of the nonlinear theory mentioned at the beginning of this section,
we conclude that there exists a chaotic layer in the vicinity of
 $E=E_{s1}$, the width of
it is determined by condition (13). Then we get 
\begin{equation}
0 < E_{s1}-E_0 < (\frac{3}{2}\omega\delta E_{s1})^{\frac 2 5} .
\end{equation}
\begin{figure}
\epsfxsize=8cm
\epsffile{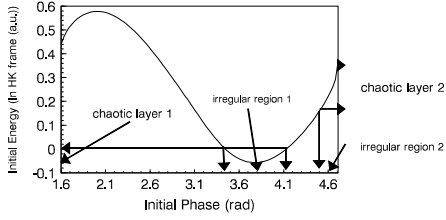}
\smallskip
\caption{
The chaotic layers predicted theoretically, showing 
coincidence with those unresolved regions. $F=0.06 a.u., \omega=0.07 a.u.$
 }
\end{figure}
In Fig.6 we plot the initial energy vs the initial field phase. The above
inequality determines a chaotic layer. We find that this chaotic  layer
coincides  with the larger one of the two unresolved regions in Fig.1.

B. Case 2, $E_{s2} = \frac{\epsilon^2}{2\omega^2}$. 
The jump of energy
can also be rewritten in the same form as in Eqs.(24) and (25) . The
$v_{c2}$ is derived   from Eq.(18) by assuming that $E_0=E_{s2}$, $X=0$.

The change of  phase is derived by evaluating the following integral,
\begin{equation}
\bar\phi=\phi+\omega\int_0^{x_0}dx/\sqrt{2E_0+2/(x^2+\alpha)^{\frac 1 2}} ,
\end{equation}
where $x_0$ is the initial position which is given by Eq.(7).
As $E_0 \to E_{s2}$, $X_0 \to \infty$ and the following approximation is applicable,
\begin{equation}
X_0 \simeq -1/2Fcos\omega t_0 ,\,\,\,\,
E_0 \simeq \frac{\epsilon^2}{2\omega^2} + 2F cos\omega t_0 .
\end{equation}
Then we obtain approximately $X_0 \simeq \frac{1}{E_0 - E_{s2}}$. As $E_0 \to
E_{s2}$, we approximate Eq.(28) by
\begin{equation}
\bar \phi = \phi + \frac{1}{\sqrt{2E_{s2}}}\frac{\omega}{E_{s2}-E_0} .
\end{equation}
Similar discussions as in case 1 can be done.  We obtain the
 following expression
which determines a chaotic layer,
\begin{equation}
0 < E_{s2}-E_0 < (\frac{\omega\delta E_{s2}}{\pi\sqrt{2E_{s2}}})^{\frac 1 2} .
\end{equation}

Figure 6 shows that this chaotic layer coincides with the smaller
one of the two unresolved regions in Fig.1. Actually, in case 2,
 the change of energy $\delta E_{s2}$ for the first collision
 is large enough
 . This fact makes
 it possible that 
 these points will fall into irregular region one finally. That is, in the
 Poincare section, the two chaotic layers expressed by Eqs. (27) and (31)
 respectively are located in the same  connected  region.
\section*{V. ATI Spectra and HOHG in the Rescattering Process }
Because the amplitude of the field varies slowly with time in the
tunneling regime 
 , the phase dependence of the ionization rate
can be expressed as follows$^{[8]}$,
\begin{equation}
\Gamma(\phi) =\frac{4}{3}\exp(-\frac{2}{3|\epsilon cos\phi|})
\,\,\,\,a.u. .
\end{equation}
In terms of the original model in which the Coulomb potential and 
rescattering are neglected, the ATI energy of an ionized electron in linearly
polarized laser beams takes the form
\begin{equation}
E_{ATI}(\phi) = 2 U_p sin^2\phi .
\end{equation}
Then the energy distribution of the ionized electrons is evaluated  by
the following expression,
\begin{equation}
P(E_{ATI}) dE_{ATI} = \Gamma(\phi) d\phi ,\,\,\,\,
P(E_{ATI}) = \Gamma(\phi) \frac{d\phi}{dE_{ATI}} .
\end{equation}
The plots of the distribution of electron energy  are shown in Fig.7,
which represents a steeply  decreasing curve.
Some ionization experiments are performed in the tunneling regime 
to test and verify  the  quasi-static theory. Analysis shows
that
qualitatively the prediction of the two-step theory agrees
with the experimental data,
and it  gives a much smaller fraction of the high-energy ionized
electrons.$^{[1]}$

In terms of the improved model, because of the rescattering effects, the
final energy of the detected electrons in a short pulse experiment can be
larger than $2 U_p$. The initial phase may be divided into four different
regions: [1.57,3.32], [3.32,4.05],[4.05,4.52] and [4.52,4.71]
 for $\omega =0.07 a.u , \epsilon = 0.06 a.u.$, and [1.57,3.30], [3.30,3.70],[3.70,4.69] and
 [4.69,4.71] for
$\omega = 0.04242 a.u. , \epsilon = 0.06 a.u.$. The trajectories initiating in different regions
show quite different behaviors (Fig.8). The electrons starting from
the first region drift away from the nucleus without a collision. Those
electrons  originating from the third region must recross the original
point once before completely escaping. However, the electrons initiating
in regions 2 and 4 can undergo multiple collisions with the nucleus
and absorb more photons in this rescattering process. As was mentioned
above,  the trajectories starting from regions 1 and region 3 are regular,
and from regions 2 and 4 may be irregular. In the situation that the
initial phase is in region 4, the tunneling barrier is too
broad  for any significant release to occur.
Then the effects of chaotic rescattering in region 2 may influence
strongly the final energy distribution.

Because of chaotic rescattering  the function $E_{ATI}$ is
not differentiable almost everywhere in those irregular regions . The usual
Riemann integral can not be applied to obtain the energy distribution, and
the Lebesgue integral is introduced to deal with the problem. The
probability of the kinetic energy of electrons falling into
the interval $[E,E+\Delta E]$ is
\begin{equation}
P = Les.\int_S E_{ATI}(\phi) \Gamma(\phi)d\phi ,
\end{equation}
where $S$ denotes a set of initial phase $\phi$ satisfying $E<E_{ATI}(\phi)
< E+\Delta E, \Delta E = U_p/10$.

\begin{figure}
\epsfxsize=8cm
\epsffile{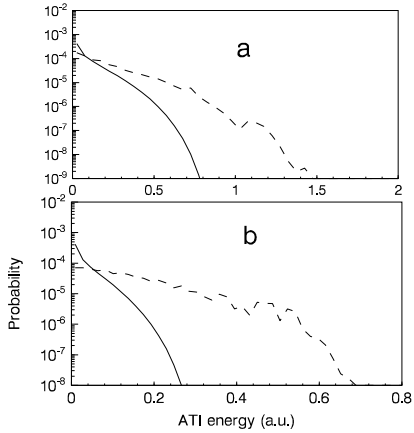}
\smallskip
\caption{
Energy spectra for $a)F=0.06 a.u.,\omega=0.04242 a.u.
\,\, b) F=0.06 a.u.,\omega=0.07 a.u.$. The solid lines and dotted lines
denote the results from the original model and the present model respectively.
 }
\end{figure}
In Fig.7 we plot the energy distribution according to the above formula.
It shows clearly that rescattering increases the fraction
of the electrons with higher energy.
This is because the electrons which  have more probability
of staying in the vicinity of the nucleus will absorb more photons.
This result is also qualitatively in agreement with that of our
quantum approaches ${^[8]}$.
In  numerical simulations, 1000 initial tryjectories are
used. We also
calculate  the ATI spectra by using 10000 trajectories . We find that
they are  agreeable except for a little fluctuation in the higher
energy range.

Another consequence of the electron-ion interaction is emission of light.
 The emission
can be calculated from the expectation value of the dipole operator
\begin{equation}
d(t) = <\psi|ex|\psi> ,
\end{equation}
where $\psi$ is the Floquet state.

In terms of the symmetry of the Hamiltonian $H_0(x,p)=H_0(-x,-p)$ and the
periodicity of the external field, Floquets theorem guarantees that
the Fourier decomposition of $d(t)$ contains neither even 
nor constant harmonic but odd harmonics dipole component.$^{[16]}$

After tunneling the density of  states available to electron is
large enough so that its subsequent evolution should be accurately
characterized by classical motion. Then the spectral analysis of
the classical trajectories can reveal the mechanism responsible
for high-order harmonic generation in the rescattering process.
We consider the power spectrum or spectral density of the dipole
moment function $x(t)$ in the classical case,
\begin{equation}
I(\omega) = (1/2\pi){lim}_{T \to \infty} (1/2T)<|\int_0^{2T}
ex(t)exp(-i\omega t)dt|^2> ,
\end{equation}
where $<>$ indicates an average over field phases $\phi$.

Actually analysis of a single trajectory is sufficient to
illustrate the underlying mechanisms.$^{[17,18]}$
Four types of classical trajectory related to rescattering are shown
in Fig.8. For regular trajectories (types 1 and 3) the power
spectrum consists of only a single peak precisely located at
the laser frequency because Rayleigh scattering is dominant in this case.
Things are much different for trajectories
originating from  in regions 2 and  4.
Figure 9 shows the dipole moment
function and its corresponding power spectrum for 
a trajectory initiated in the irregular regions.
In this case , the electron is first pumped by the external fields
far away from the nucleus for a long time before  returning to the ion
and emits light. Only  when the electron is close to the nucleus,
 can the electron exchange energy efficiently with the electromagnetic
 field.   After oscillating around the nucleus for about 10 optical
 cycles, the electron absorbs more photons and is ejected to the continuum.
 A power spectrum analysis of the dipole moment is made in the
 time interval $[10000,11000]$ when the electron oscillates near the
  nucleus. The power spectrum shows that high harmonics up to 33rd are
  generated in the process.
  The magnitude of the peaks decreases  monotonically approximately
  with increasing order. The peak corresponding to 33rd harmonic
  is about three orders lower than that of Rayleigh scattering.
  The even harmonics component appearing in the power spectrum
  can be removed by averaging over an ensemble
  of classical trajectories.
\begin{figure}
\epsfxsize=8cm
\epsffile{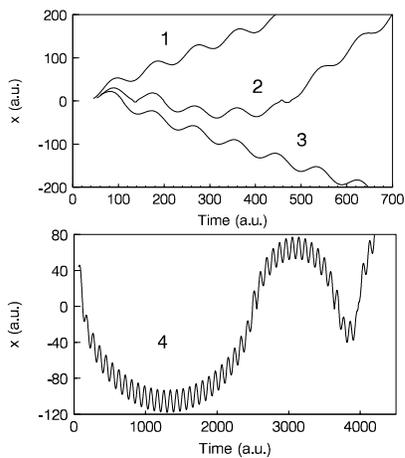}
\smallskip
\caption{
Typical trajectories initiated in four different regions.
 }
\end{figure}
\begin{figure}
\epsfxsize=8cm
\epsffile{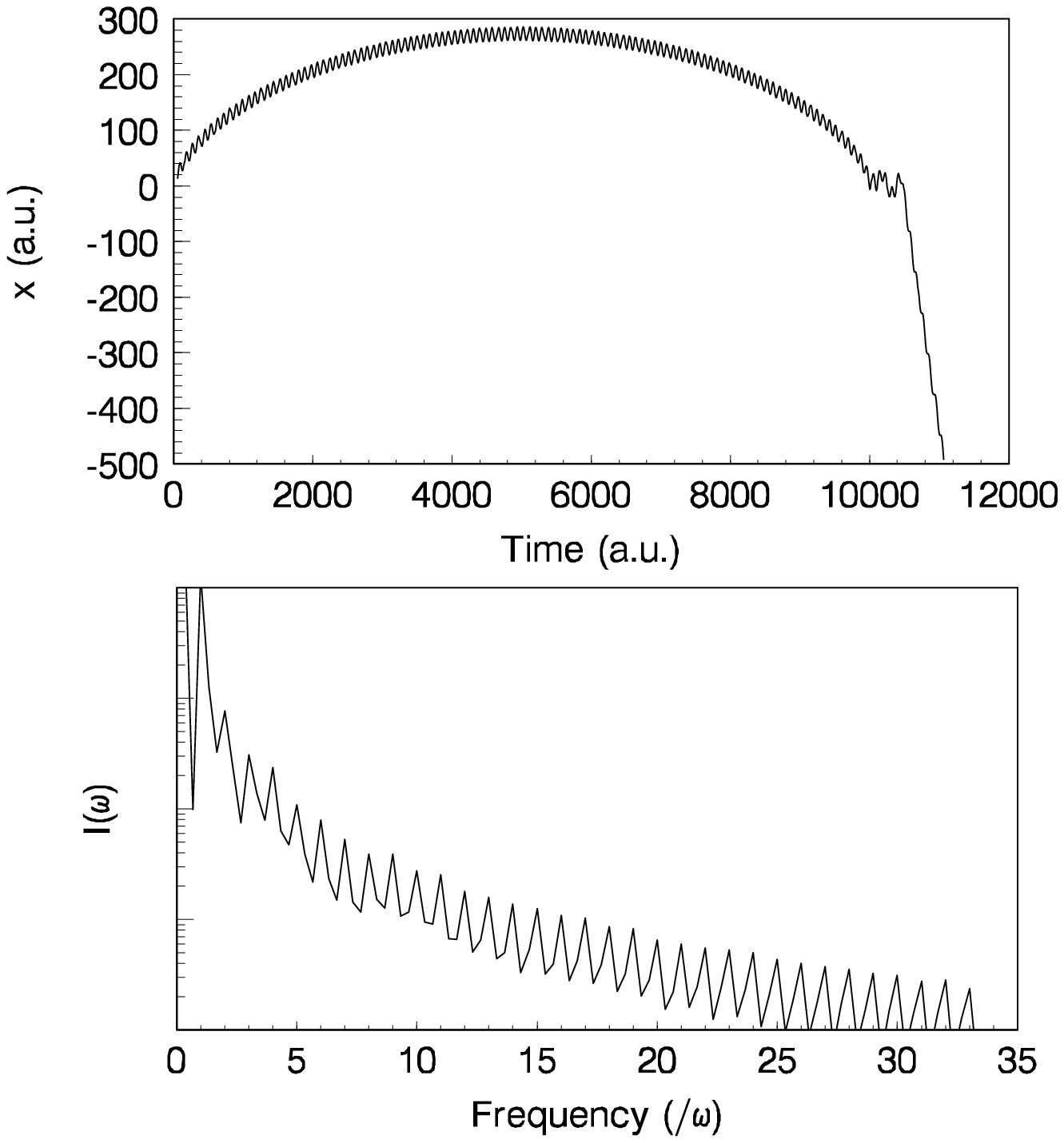}
\smallskip
\caption{
The dipole moment  as a function of time
 and corresponding power spectrum for a typical trajectory
  initiated in irregular regions.
 $\phi_0 = 3.3250968504$, 
 $F=0.06 a.u.,\omega=0.07 a.u.$.
 }
\end{figure}

  In the high-intensity regime  a new contribution to
  harmonic emission becomes possible. That is , while an electron
  is near the nucleus it can make a transition back to the ground
  state, emitting a high-energy photon. Obviously the energy of
  the resulting photon corresponds to the energy (return
   energy $E_r$) the electron   has  when it collides with the core.
   This return energy depends strongly on the field  phase
   at which the electron escapes through the potential barrier.
   In Figs.10(a) and 11(a) we plot the return energy vs the phase for
   $\omega=0.04242 a.u.$ and $\omega=0.021 a.u.$ respectively. The 1000 initial
   phases distributed equally in the interval $[\pi/2,3\pi/2]$ are
   used in the calculation. In Figs.10(b)
   and 11(b) we show  the histograms of the return energies weighted
   by the tunneling rate appropriate to the initial conditions
   for each trajectory. The result is a broad, flat
   distribution followed by an abrupt cutoff. The sharp cutoff in
   the electron energy occurs at $3.5U_p$ for $\omega = 0.04242 a.u.$ and
   at $3.3U_p$ for $\omega = 0.021 a.u.$ respectively. We think this
   is the physical origin of the cutoff law for high harmonic
   radiation.$^{[6,19]}$ The smaller the field frequency,
   the more  the tunneling
   condition is satisfied. Therefore the value of the cutoff energy
   tends to $3.2U_p$, which has been supported by experimental
   evidence.
\begin{figure}
\epsfxsize=8cm
\epsffile{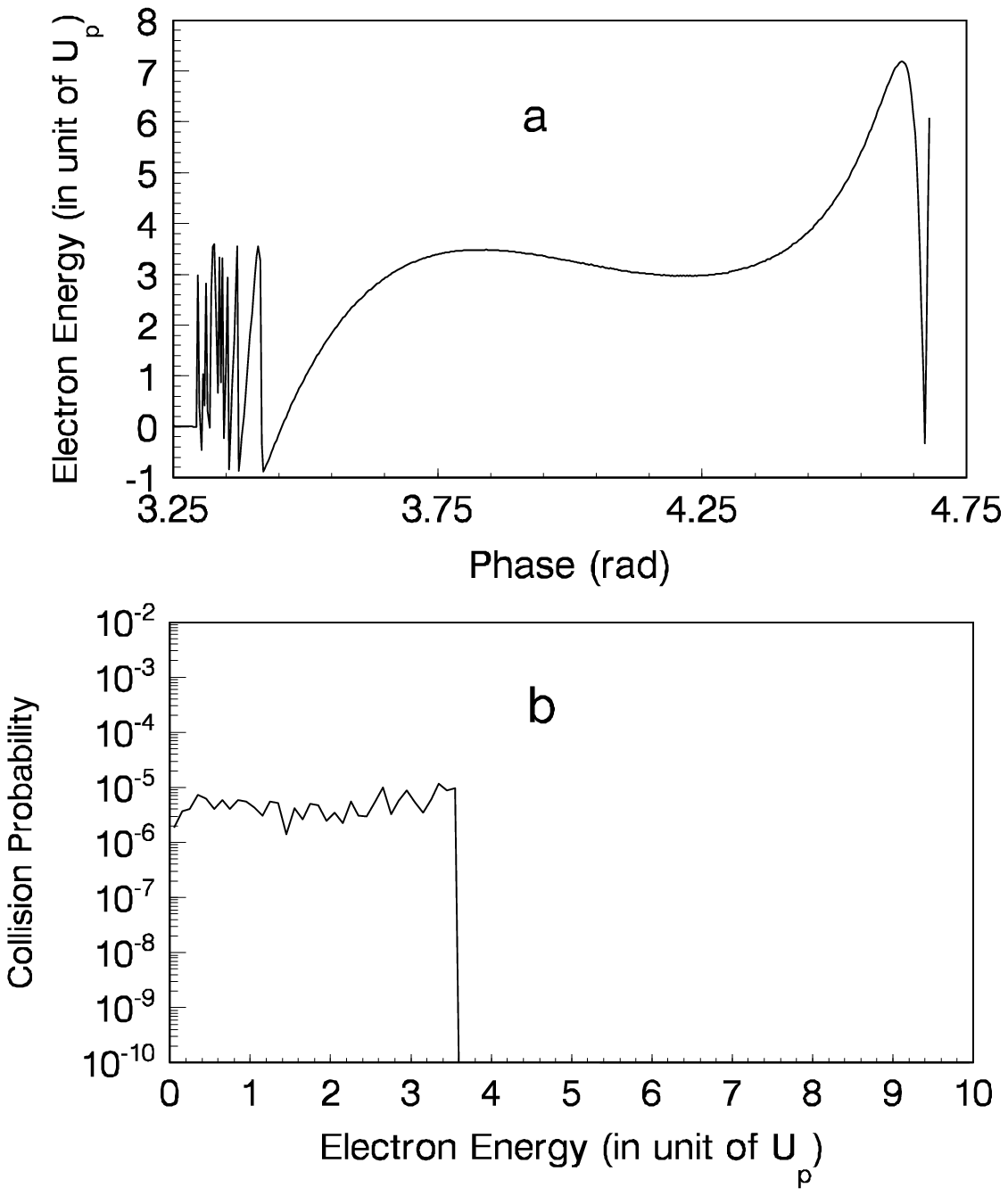}
\smallskip
\caption{
a) The energy of returning electron  as a function of
field phase. b) Energy histogram for electron trajectories returning
to the nucleus. $\omega=0.04242 a.u.$ and $F=0.06 a.u.$.
 }
\end{figure}
\begin{figure}
\epsfxsize=8cm
\epsffile{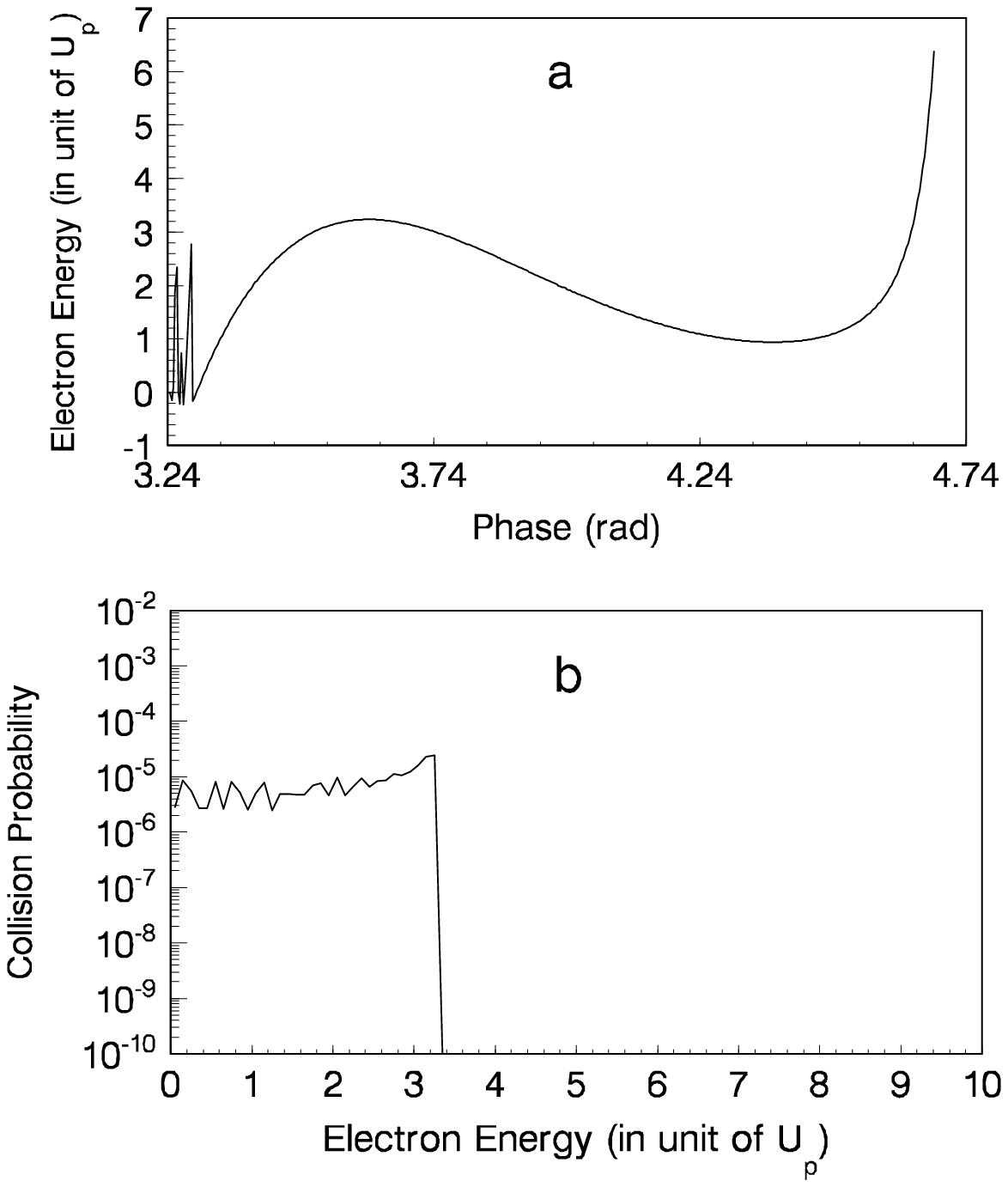}
\smallskip
\caption{
a) The energy of returning electron  as a function of
field phase. b) Energy histogram for electron trajectories returning
to the nucleus. $\omega=0.021 a.u.$ and $F=0.06 a.u.$.
 }
\end{figure}

   \section*{VI. Dynamical Structure in Poincare Section}
   Irregular behavior of the electrons in the rescattering  process is
   characterized by the singularity of the final ATI energy distribution.
   This singularity is attributed to chaotic scattering in system (15).
   The system studied in  previous sections constitutes an excellent
   physical system for the  investigation of chaotic scattering.
   The physical  process that we have in mind involves the rescattering
   in the OFI problem.

   To study chaotic scattering, we choose the initial condition
   as the points in phase space well in the asymptotic region of the
   potential, typically, $X=100$ $a.u.$ and $P$ in the range $[0,-1.5]$ $a.u.$.
   Figure 12 shows the stroboscopic map, ${(X(t),P(t))|_{\omega t = 0}
  (mod 2\pi)}$, for such an ensemble of 200 scattering trajectories. Further
  the system can flow arbitrarily between the bounded and unbounded regimes,
  so it is a non-compact system. The phase section shows a rich and complicated
  structure. As the system is periodically  perturbed, the Poincare section
  is naturally defined as ${\sum:\omega t=2k\pi}$. The corresponding
  Poincare mapping is expressed as
  \begin{equation}
  M: (X',P') = M\circ (X,P) = (F(X,P),G(X,P)) ,
  \end{equation}
  where $(X',P')$ is the point at $\omega t=\omega t_0 + 2\pi$ of the
  trajectory initiating  in $(X,P)$ at $\omega t_0$.
  Its corresponding tangent mapping will be
  \begin{equation}
  TM:(\delta X',\delta P') =(\frac{\partial F}{\partial X}\delta X +
                    \frac{\partial F}{\partial P}\delta P,
                    \frac{\partial G}{\partial X}\delta X +
                    \frac{\partial G}{\partial P}\delta P ) .
\end{equation}
  In the following
  discussions  $ t_0$ is set to be zero  without losing
   generality.

  The varitional equations  of the Hamiltonian system are
 
$$  \frac{d(\delta X)}{dt} = \frac{\partial^2 K}{\partial P\partial X}\delta X
                 +\frac{\partial^2 K}{\partial P\partial P}\delta P , $$
   \begin{equation}
  \frac{d(\delta P)}{dt} = -\frac{\partial^2 K}{\partial X\partial X}\delta X
                 -\frac{\partial^2 K}{\partial X\partial P}\delta P .
  \end{equation}

  Equations (16) and (40) constitute a four-dimensional non-autonomous
   dynamical system. Tracing the trajectories initiating in
   $(X_0,P_0, 1,0), t_0=0$ and $(X_0,P_0, 0,1), t_0=0$  by using
    a $4{th}$ order Runge-Kutta algorithm in $x$ and $p$,
     we can get the solutions
   $(X_1,P_1, \delta X_1,\delta P_1)$ and $(X_1,P_1, \delta X_2,\delta P_2)$
   at time
   $\omega t=2\pi$ respectively . The Pioncare mapping and
   its tangent mapping can be
   calculated  numerically according to following expressions:
   \begin{equation}
   X_1 = F(X_0,P_0), P_1 = G(X_0,P_0) , 
   \end{equation}
$$\frac{\partial X_1}{\partial X_0} =
\frac{\partial F}{\partial X_0}=\delta X_1 ,$$
\begin{equation}
\frac{\partial P_1}{\partial X_0} =
\frac{\partial G}{\partial X_0}=\delta P_1 .
\end{equation}
$$\frac{\partial X_1}{\partial P_0} =
\frac{\partial F}{\partial P_0}=\delta X_2 ,$$
\begin{equation}
\frac{\partial P_1}{\partial P_0} =
\frac{\partial G}{\partial P_0}=\delta P_2 .
\end{equation}
\begin{figure}
\epsfxsize=8cm
\epsffile{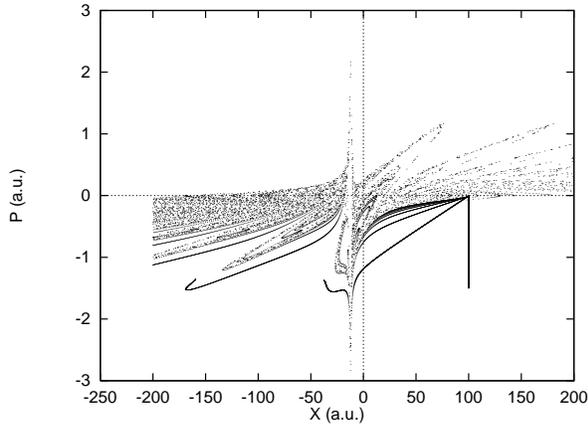}
\smallskip
\caption{
Stroboscopic map for 200 trajectories originating
asymptotically at $X=100$ with $P$ in range $[0,-1.5]$.
$F=0.06 a.u.,
\omega = 0.07 a.u.$.
 }
\end{figure}
\begin{figure}
\epsfxsize=8cm
\epsffile{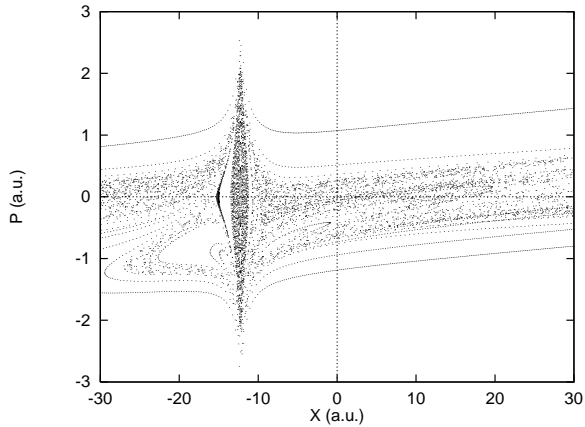}
\smallskip
\caption{
 Phase plane of Poincare mapping $M$ for
$F=0.06 a.u.,
\omega = 0.07 a.u.$.
 }
\end{figure}
 Since the Cartan differential 2-form is
 conserved under Hamiltonian flow, the mapping $M$ should
 conserve $dX\wedge dP$, that is , it is area-preserving.
 Because of  the symmetry of the Hamiltonian system, the 1-periodic
 fixed points are located on the $X$-axis. The Newton-Raphson algorithm is
 employed to locate these fixed points precisely. They are
 $P_1(-15.95186600603247,0)$,
 $\,\, P_2(-15.02,0)$,
 $\,\, P_3(-12.25,0)$,
 $ \,\,P_4(-9.30,0)$,
 $ \,\,P_5(2.80,0)$,
 $ \,\, P_6(15.87,0)$. $P_2$ and $P_3$ are stable elliptic
 fixed points, others are unstable hyperbolic fixed points. Figure 13
 shows the phase plane of the mapping $M$. Around $P_3$ there is
  a larger stable
 region where KAM curves exist. The stable region around  $P_2$ is too
 small to be distinguished in the plots.

 From the symmetry  $K(X,-P,-t) = K(X,P,t)$, the
 stable manifolds and unstable manifolds are symmetric about the X-axis.
 In Fig.14 we plot the unstable manifolds of these four hyperbolic
 fixed points by using the method suggested in Ref.$[21]$.
 Clearly , they intersect the X-axis. So we conclude that
 the unstable manifolds and stable manifolds intersect
  each other transversely. The
 intersection points constitute a hyperbolic invariant set with
 a fractal dimension which has been mentioned in section III. The Smale
structure occurs in the phase plane. The is the dynamical source of
  chaotic scattering.
 Comparing Figs.12 ,13 and 14, we find that the unstable manifolds   organize
 the scattering pattern and the phase plane structure.
\begin{figure}
\epsfxsize=8cm
\epsffile{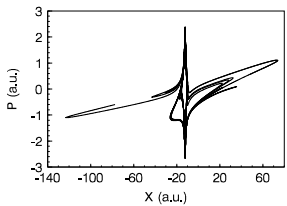}
\smallskip
\caption{
The unstable manifolds of period-1 saddles of mapping $M$ for
$F=0.06 a.u. , \omega = 0.07 a.u.$.
 }
\end{figure}

 \section*{VII. Discussions and Conclusions}
Recent advances in classical nonlinear dynamics and chaos
 have had important applications in the
 description of the photoabsorption spectrum of Rydberg atoms in strong
 magnetic fields,$^{[22]}$  and the microwave ionization of
 highly excited hydrogen atoms.$^{[23]}$ These discussions extremely
 enrich the atomic physics. More recently,
 some pioneer nonlinear dynamicist begin to
 deal with the topics of atoms in intense fields. Sundaram and
 Jensen$^{[24]}$ connect the ionization suppression with 'scarred' wave
 functions. Benvenuto {\it et al.}$^{[25]}$ estimate the parameter
 regime of atom stabilization  according to the border of chaos.
 Cocke and Reichl$^{[26]}$ calculate the HONG in the frame of
 1D model atom and conclude that it implies chaos.
 Richards$^{[27]}$ has studied 1D excited  hydrogen atoms in
 short-pulsed high-frequency fields and find the signs of incipient
 chaos.
 The purpose of this paper is to apply the  nonlinear theory in
 ATI and HOHG according to a 1D description of  
 optical field ionization in the regime of low field frequency.
 We discover a new physical process - chaotic scattering underlying
the optical field ionization process.
  Modern nonlinear theory is applied to analyze the chaotic behavior
  of this nonlinear system. Chaotic layers and self-similar fractal
   structure are found  in this model. Otherwise we find that
   unstable and stable manifolds
   organize the chaotic scattering pattern.

   In another aspect, the above-threshold ionization energy
   spectra and high-order harmonic generation are discussed
   according to the improved quasi-static model. The rescattering
    effects on ATI spectra and high-order harmonic generation
    are discussed.
    The result that rescattering  increases the fraction of the
    electrons with higher energy is qualitatively in
    agreement with our previous works from a quantum approach.
    The cutoff law for high harmonic radiation is identified
    by our new model.   In particular,
    We  have shown the connection between
    regular and irregular trajectories and ATI , HOHG phenomena.
    We find that chaotic behavior plays an important role in
    these processes.

    In this paper, our discussions are restricted to 1D
    model . A question that can not be evaded is: Whether the above
    discussions are helpful to the understanding of  the ionization process
    of a real atom in a three-dimensional (3D) space ? The Key point of this
    problem is whether multiple collisions ( resulting in chaos) still
    occur in a 3D case. It seems that there are only
    trivial number of multiple return trajectories. Actually, this
    is a specious argument. In our 3D extension of
    the above model, numerical simulations show that there do exist
    a number of (not trivial number) trajectories which return to the
    core more than once and those multiple collisions trajectories (chaos)
    play an important role in energy spectra and angular distribution
    of the final hot electrons (results in details will be reported elsewhere)
    . Therefore, we believe that our discussions on 1D
    model is much helpful to the understanding of  the ionization process of a real
    atom in a 3D case.

    Notes: After we completed this manuscript, we find that the authors
    of Ref. $[28]$ have made some researches on the effects of multiple
    returns on double ionization for 3D model, and show an
    enhancement by more than an order of magnitude.

\newpage
\section*{Captions of Figures}
\begin{itemize} 
\item Fig.1 Plots of electron energy  vs field phase. Dashed lines and
solid lines denote the results from original model
and the  present model respectively.
a) $F=0.06 a.u. ,\omega=0.07 a.u. \,\,$b)$F=0.06 a.u.,\omega=0.04242 a.u.$.
\item Fig.2 Trapping time vs field phase. $F=0.06 a.u.,\omega=0.04242 a.u.$.
\item Fig.3 Successive magnification of the plot for $F=0.06 a.u., \omega=0.07 a.u.$.
\item Fig.4 Calculation of fractal dimension
in the case $F=0.06 a.u., \omega=0.07 a.u.$
\item Fig.5 Probability of multiple collisions obtained from making statistics on
4000 trajectories corresponding to different initial phase (circle). The
straight line denotes the scaling law mentioned in the text.
\item Fig.6 The chaotic layers predicted theoretically, showing 
coincidence with those unresolved regions. $F=0.06 a.u., \omega=0.07 a.u.$
\item Fig.7 Energy spectra for $a)F=0.06 a.u.,\omega=0.04242 a.u.
\,\, b) F=0.06 a.u.,\omega=0.07 a.u.$. The solid lines and dotted lines
denote the results from the original model and the present model respectively.

\item Fig.8 Typical trajectories initiated in four different regions.
\item Fig.9 The dipole moment  as a function of time
 and corresponding power spectrum for a typical trajectory
  initiated in irregular regions.
 $\phi_0 = 3.3250968504$, 
 $F=0.06 a.u.,\omega=0.07 a.u.$.
\item Fig.10 a) The energy of returning electron  as a function of
field phase. b) Energy histogram for electron trajectories returning
to the nucleus. $\omega=0.04242 a.u.$ and $F=0.06 a.u.$.
\item Fig.11 a) The energy of returning electron  as a function of
field phase. b) Energy histogram for electron trajectories returning
to the nucleus. $\omega=0.021 a.u.$ and $F=0.06 a.u.$.
\item Fig.12 Stroboscopic map for 200 trajectories originating
asymptotically at $X=100$ with $P$ in range $[0,-1.5]$.
$F=0.06 a.u.,
\omega = 0.07 a.u.$.
\item Fig.13 Phase plane of Poincare mapping $M$ for
$F=0.06 a.u.,
\omega = 0.07 a.u.$.
\item Fig.14 The unstable manifolds of period-1 saddles of mapping $M$ for
$F=0.06 a.u. , \omega = 0.07 a.u.$.

\end{itemize}

\end{document}